\documentclass[aps,prl,twocolumn,amsmath,amssymb,showpacs]{revtex4}
\usepackage{graphicx}% Include figure files
\usepackage{dcolumn}% Align table columns on decimal point
\usepackage{bm}% bold math
\usepackage{psfrag} % pour modifier des donnees d'un ps
\usepackage{color}
\begin{document}
\title{Properties of holons in the Quantum Dimer Model}
\author{Didier Poilblanc${^1}$}
\affiliation{ ${^1}$ Laboratoire de Physique Th\'eorique, CNRS \&
Universit\'e de Toulouse, F-31062 Toulouse, France 
}
\date{\today}
\begin{abstract}
I introduce a doped two-dimensional quantum dimer model 
describing a doped Mott insulator and retaining the original Fermi 
statistics of the electrons. This model shows a rich phase diagram including
a d-wave hole-pair unconventional superconductor 
at small enough doping and a bosonic superfluid at large doping.
The hole kinetic
energy is shown to favor binding of topological defects to the bare fermionic
holons turning them into bosons, in agreement with arguments 
based on RVB wave-functions. Results are discussed in the 
context of cuprates superconductors.

\end{abstract}

\maketitle
%
%------------------------------------------------------------------------------
% Paper
%------------------------------------------------------------------------------
%

\begin{figure}[htpb]
\begin{center}
\includegraphics[width=0.35\textwidth,clip]{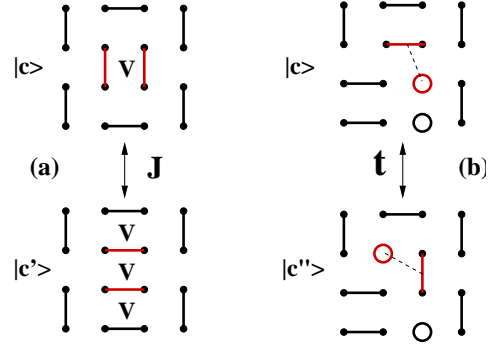}
\caption{ (color online) 
Dimer flip (a) and holon hopping (b) in the QDM.}
 \label{configs}
 \end{center}
 \end{figure}

The discovery of high temperature superconductivity in copper oxydes 
triggered unprecedented efforts to understand the role
of strong correlations in electronic systems. In Anderson's
milestone paper~\cite{anderson}, the "mother"
correlated insulator, on top of which superconductivity emerges
under doping, was proposed as a Resonating Valence Bond (RVB)
state where localized electrons are paired up into singlet (bond) dimers due to antiferromagnetic (AF) exchange.
Here, a substantial energy is gained from quantum resonance 
between different
(short-range) dimer coverings. 
Re-formulating these ideas~\cite{note_Sp2N} at the Hamiltonian level,
Rokhsar and Kivelson constructed 
a Quantum Dimer Model (QDM) involving orthogonal dimer coverings
and local dimer flips~\cite{rokhsar} (see Fig.~\ref{configs}(a)), 
aimed to capture the physics of systems that possess a spin (pseudo-) gap.
Surprisingly, the QDM proved also to be
relevant in a variety of distinct fields such as e.g. frustrated 
Ising models~\cite{moessner}, spin-orbital models~\cite{vernay}
or superconducting junction arrays~\cite{albuquerque}.
Although the ground state (GS) of the QDM on the square lattice 
is a Valence Bond Crystal (VBC) breaking lattice symmetry~\cite{syljuasen}, 
finite doping is expected to melt the crystal.
In that respect, the doped QDM is of great interest since potentially more tractable 
than microscopic t--J or Hubbard models while effectively retaining their basic 
physical processes (spin exchange and hole hopping).

\begin{figure}[htpb]
\begin{center}
\includegraphics[width=0.45\textwidth,clip]{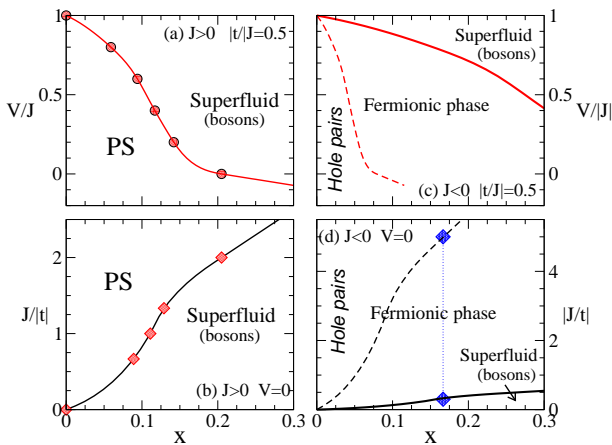}
\caption{ (color online) 
 Phase diagram of the doped QDM's versus doping ($x$) and
 $V/|J|$ (a,c) or $|t/J|$ (b,d). The Frobenius (a,b) and non-Frobenius (c,d) 
cases labelled on the plots as $J>0$ and $J<0$, respectively, 
 (see text) are compared. The $x=0$ insulator has VBC order.
The symbols in (a,b) [resp. (d)] are obtained from the data of
Fig.~\protect\ref{Maxwell_bosons} [resp. Fig.~\ref{OP_Sign}].
}
 \label{Phase_diag}
 \end{center}
 \end{figure}

The exact nature of holons, 
the charged spinless excitations of the doped RVB state, has
been debated over the last two decades. In particular, their statistics
is still unclear. 
Earlier work based on a variational RVB 
wavefunction~\cite{statistics_read}
suggested that hole excitations are (weakly interacting) {\it fermions}.  
Such a conclusion, although only justified for small enough kinetic energy,
was reproduced in the context of the doped 
QDM~\cite{statistics_kivelson}.
Furthermore, it was argued~\cite{statistics_kivelson} that the holon 
statistics should in fact be dictated by energetics considerations
and that, under some circonstancies, a holon could bind to a ``vortex'',
(see e.g. Ref.~\onlinecite{statistics_read} for a simple definition) 
leading to a {\it bosonic} composite. 
$Z_2$ gauge theories have recently brought
powerful new tools~\cite{Fisher} to describe such phenomena. 
However, a more quantitative large-scale numerical investigation
of the microscopic doped QDM is clearly needed.
In this Letter, I perform such a program using Exact Diagonalizations 
(ED) of periodic $6\times 6$ and $8\times 8$ clusters~\cite{note_ED} and propose a
semi-quantitative phase diagram. In contrast to previous investigations of doped QDM's~\cite{Frobenius_doped_QDM} with non-positive off-diagonal
matrix elements (named ``Frobenius'' QDM's), I consider here
the {\it non-Frobenius} doped QDM which incorporates the Fermi 
statistics~\cite{sign_problem} of the original electrons.
Major differences in the phase diagrams shown in Fig.~\ref{Phase_diag}
are observed and discussed in the text. Of particular interest are:
(a) a new unconventional {\it d-wave} superconducting phase (with possible lattice symmetry 
breaking) and (ii) a spectacular change of the hole statistics, 
from Fermi to Bose, by increasing doping or kinetic energy.

% Figure 0 (Hamiltonian) 

{\it The non-Frobenius doped QDM model.}---Following 
Ref.~[\onlinecite{rokhsar}] and recent 
investigations~\cite{Frobenius_doped_QDM}, I start 
with the quantum hard-core dimer-gas on the two-dimensional 
square lattice defined by the Hamiltonian:
\begin{eqnarray}\label{hamilt}
H &=& V \sum_{c} N_c | c \rangle \langle c | -J \sum_{(c,c')} | c' \rangle
\langle c | -t \sum_{(c,c'')} | c'' \rangle \langle c | 
\nonumber
\end{eqnarray}
where the sum on $(c)$ runs over all dimer coverings (containing a fixed 
amount of $N_h$ vacant sites), 
$N_c$ is the number of flippable plaquettes, the sum on $(c',c)$ runs over all
configurations $| c \rangle$ and $| c' \rangle$ that differ by a single
plaquette dimer flip, and
the sum on $(c'',c)$ runs over all configurations $| c \rangle$ and $|
c''\rangle$ that differ by a single hole (or vacant site) hopping 
along a plaquette diagonal as pictured in Fig.~\ref{configs}.
In this formulation, {\it bare} holons
(i.e the moving vacancies) have Bose statistics. Since the system 
is originally composed of electrons of Fermi statistics, 
a faithful description should then assume that 
dimer configurations (with vacancies) are created by sets of  
(spatially symmetric) dimer operators expressed in the {\it fermionic} 
representation as e.g. in 
Refs.~[\onlinecite{statistics_kivelson,note_Sp2N}], 
i.e. $d_{ij}^\dagger=\frac{1}{\sqrt{2}}(f_{i\uparrow}^\dagger f_{j\downarrow}^\dagger
+f_{j\uparrow}^\dagger f_{i\downarrow}^\dagger)$ where $f_{i\sigma}^\dagger $ creates 
an electron of spin $\sigma$ and $i$ and $j$ are neighboring sites.
The original AF bond couplings  
yield then, in the QDM language, $J<0$ and hence a {\it non-Frobenius} Hamiltonian with 
positive $-J=|J|$ off-diagonal matrix elements. 
Note that a bosonic convention for the dimers can equally well be used as will be discussed
later on. Sofar the 
sign of $t$ is not specified~\cite{note_sign_t}.
Phase diagrams (Fig.~\ref{Phase_diag}) are investigated 
as a function of $V/|J|$ and $|t/J|$.

% Figure 1 (Phase diagrams) 

\begin{figure}[htbp]
\begin{center}
\includegraphics[width=0.45\textwidth,clip]{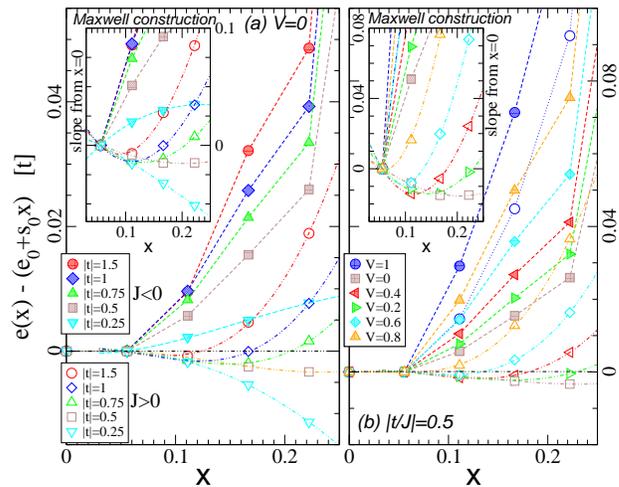}
\caption{(color online) 
On-site GS energy (per site) vs doping calculated on a $6\times 6$
cluster. Full (open) symbols are used for the
non-Frobenius (Frobenius) QDM labelled as $J<0$ ($J>0$). The linear
behavior at $x\rightarrow 0$ has been subtracted  for convenience. (a) $V=0$ and different values of $|t|$; (b) $|t/J|=0.5$ and
different values of $V$ (all in units of $|J|=1$).
Insets: mean-slopes between $x=0$ and doping $x$ plotted vs $x$. 
For $J>0$, a polynomial fit gives a minimum at $x_c$ providing the
range $[0,x_c]$ of PS (Maxwell construction) reported in
Fig.~\protect\ref{Phase_diag}(a,b). For $J<0$ no minimum at finite $x$ is seen.
} 
\label{Maxwell_bosons}
\end{center}
\end{figure}

{\it About Phase Separation}---On finite clusters, Phase Separation (PS) is signaled
by a negative compressibility i.e. the local curvature of the energy vs doping $e(x)$ curve. 
In that case, the equilibrium hole densities of the
two-component mixture can be obtained via a standard {\it Maxwell construction} as for an 
ordinary liquid-gas first-order transition. Using such a procedure
one shows that the Frobenius QDM (i.e. with $J>0$) 
phase separates at small doping~\cite{Frobenius_doped_QDM}.
The ED results for both signs of $J$ are compared in Fig.~\ref{Maxwell_bosons}; while 
for $J>0$ the $x\rightarrow 0$  curvature 
of $e(x)$ is always negative, it is always positive for $J<0$, even for small $|t/J|$ ratios. This shows that the two models behave quite differently in the immediate vicinity of the $x=0$ axis as can already 
be noticed on their respective phase diagrams (Fig.~\ref{Phase_diag})~: (i) PS for $J>0$ (consistently with 
the data of Ref.~\onlinecite{Frobenius_doped_QDM} obtained for smaller $|t/J|$ and closer to the RK point $V/J=1$) and (ii) a homogeneous phase (to be identified next) in the non-Frobenius case.
In the latter case of most interest here, I find a very linear behavior
of $e(x)$ at finite doping evolving into a region of negative curvature
as soon as $|t/J|>0.5$ suggesting that some form of PS might occur 
at {\it finite} doping. 
 
% Figure 2

\begin{figure}[htpb]
\begin{center}
\includegraphics[width=0.45\textwidth,clip]{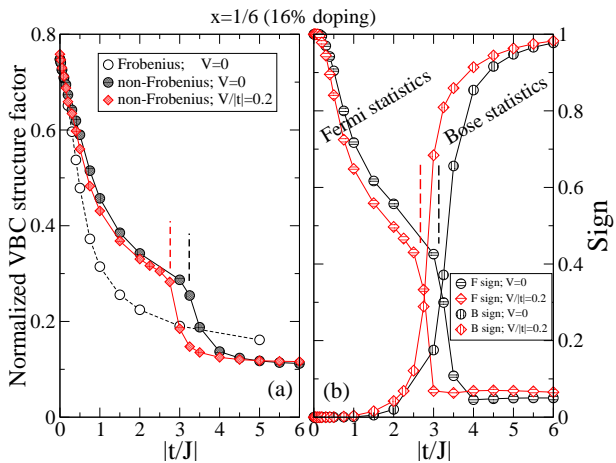}
\caption{(color online) 
(a) Columnar (i.e. ${\bf q}=(\pi,0)$) VBC structure factor (normalized to 
its $x=0$ value) versus $|t/J|$. 
(b) "Fermionic'' and "bosonic'' signs (as defined in text) vs $|t/J|$.
(a) and (b) correspond to the non-Frobenius doped QDM with $V=0$ (circles) and 
$V/|t|=0.2$ (lozenges) at $x=1/6$ (computed on a $6\times 6$ cluster). 
In (a) additional data for
the Frobenius case are shown for comparison.
Vertical dotted segments give the 
approximate separations between the phases discussed in the text.}
\label{OP_Sign}  
\end{center}
\end{figure}

{\it Melting of VBC order}--- How does the $x=0$
VBC order evolve under finite doping is of great interest. Although,
unfortunately, no finite-size scaling is possible here~\cite{sign_problem}, a comparison with the Frobenius model (for which large systems could be studied, see~\cite{Frobenius_doped_QDM}) is very instructive.
As shown in Fig.~\ref{OP_Sign}(a), the relevant ${\bf q}=(\pi,0)$
dimer-dimer correlator, decreases rapidly with the hole kinetic energy. However, at small enough $|t/J|$, the magnitude of the 
structure factor for the non-Frobenius model is always, on our clusters, larger than the same quantity in the Frobenius model.
Since the later was shown to
remain finite in the thermodynamic limit~\cite{Frobenius_doped_QDM},
this strongly suggests that there is, also in the non-Frobenius model,
a finite region in the vicinity of 
$x=0$ where VBC order survives. Note however that this phase is
{\it not} phase-separated as in the Frobenius case.
Interestingly enough, Fig.~\ref{OP_Sign}(a)
also reveals at larger $|t/J|$ a sudden drop of the structure factor
that we shall attribute later on to a real transition. 

\begin{figure}[htbp]
\begin{center}
\includegraphics[width=0.45\textwidth,clip]{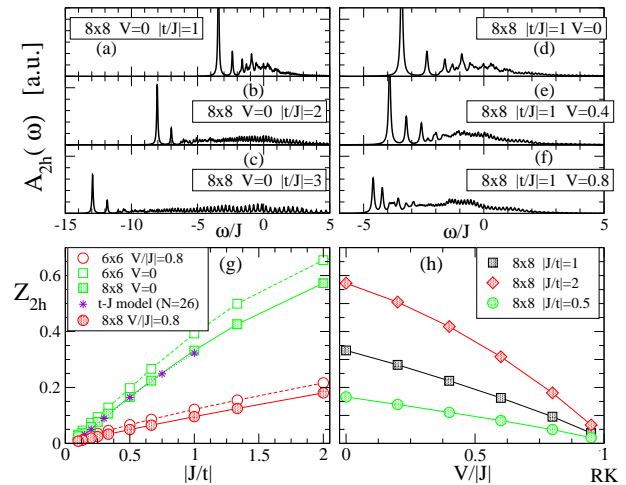}
\caption{(color online) 
(a-f) Spectral functions associated with the propagation of a 
${\bf q}=(0,0)$ d-wave pair of holes
versus frequency $\omega$ (computed on a periodic $8\times 8$ cluster) for 
fixed $V=0$ (left) or $|t/J|=1$ (right). 
The weight of the hole-pair QP peak (at the lowest energy edge) is
plotted vs $|J/t|$ (g) and $V/|J|$ (h). Comparison with 
data on a $6\times 6$ cluster in (g) suggests weak finite size effects. 
Data for the t-J model (stars) from Ref.~\protect\onlinecite{Z2h_tJ}
are also shown for comparison in (g). 
Note that 
''string resonances'' can be seen
above the QP peak in (a-f).}
\label{Pairing}  
\end{center}
\end{figure}

{\it The d-wave superconductor}--- Let us now refine the characterization of the
low doping phase in the non-Frobenius model. 
Two doped holes~\cite{note_Nh_eq2}  are found to form a zero-momentum bound-state, of 
$d_{x^2-y^2}$ symmetry providing
one chooses $t>0$~\cite{note_sign_t}. This is
revealed by the ``quasi-particle'' (QP) peak at the bottom of
the two-hole spectral functions plotted in Fig.~\ref{Pairing}(a-f). 
The associated weight $Z_{2h}$ vanishes when approaching the RK point, i.e. $V/|J|\rightarrow 1$, 
where VBC correlations become algebraic (Fig.~\ref{Pairing}(h)).
This suggests that pairing is due to confinement by the VBC, at least at large length-scales.
Incidently, a few higher energy peaks with approximate $(\frac{V}{|J|}-1)^{2/3}$
energy dependance can be associated to so-called ``string resonances'', typical of 
confinement phenomena. Interestingly enough, I also observed that the $|t/J|$-dependance of $Z_{2h}$ 
(Fig.~\ref{Pairing}(g)), $Z_{2h}\sim |J/t|^\eta$, $\eta\le 1$, is quite similar to the case of 
a two-hole pair propagating in a quantum AF~\cite{Z2h_tJ}.
Since PS does not occur in the vicinity of the $x=0$ axis one then expects a d-wave hole-pair superconductor
in some finite doping range (see Fig.~\ref{Phase_diag}(c,d)). One could even argue (see above) that, at sufficiently small
doping, d-wave (pair) superconducting and VBC orders are likely to coexist, although VBC order is not required for superconductivity. 
%providing a close analogy with bosonic {\it supersolids}~\cite{supersolids}.

% Figure 4

{\it Holon statistics: Fermi vs Bose}---Before turning to the investigation of the {\it actual} statistics of the holons, it is useful to mention briefly an alternative convention
for the non-Frobenius QDM.  
Indeed, dimer operators can be represented in the usual bosonic representation 
$d_{i\rightarrow j}^\dagger=\frac{1}{\sqrt{2}}(b_{i\uparrow}^\dagger b_{j\downarrow}^\dagger
- b_{i\downarrow}^\dagger b_{j\uparrow}^\dagger) $ in terms
of boson creation operator $b_{i\sigma}^\dagger $ carrying spin $\sigma=\pm \frac{1}{2}$, hence leading to a change of
the sign of $J$, i.e $J>0$ (provided $d_{i\rightarrow j}^\dagger$
is oriented e.g. from the A to the B sublattice). 
However, to preserve the fermionic character of the original electron, the {\it bare} holon should be 
a fermion, meaning that one now keeps track of some fixed (arbitrary) ordering of the holons~\cite{note_equivalence}.
%This mapping can be formally realized by a generalized 
%Jordan-Wigner transformation on the holons and it 
Obviously, the {\it actual} statistics of the holon
is independent on the representation used.

% Figure 3
Motivated by~\cite{statistics_read} I now define ``fermionic'' (R=F) and ``bosonic'' (R=B) average signs as:
\begin{eqnarray}\label{sign}
{\rm Sign}_{\rm R}=\sum_{\cal H_\alpha}|\sum_{c\in{\cal H_\alpha}}
\langle\Psi_0^{\rm R}| c \rangle 
\,| \langle\Psi_0^{\rm R}| c \rangle | \,|
/\sum_{c}|\langle\Psi_0^{\rm R}| c \rangle|^2\,  ,
\nonumber
\end{eqnarray}
where the first sum is performed over all classes $\cal H_\alpha$
of configurations with fixed hole positions, the second sum runs over all
dimer configurations within each class. The index ``R'' refers to 
one of the two (equivalent) representations of the non-Frobenius model (as discussed above) i.e. 
choosing {\it bare} holons with either Fermi (R=F) or Bose (R=B) statistics
together with $J>0$ (R=F) or $J<0$ (R=B).
Therefore, if holons truly behave as fermions (resp. holons) one expects 
${\rm Sign}_F\simeq 1$ (resp. ${\rm Sign}_B\simeq 1$).
Since a fermion (resp. boson) can be seen as a {\it bare} bosonic (resp. 
fermionic) holon
bound to a vortex (or fluxoid)~\cite{statistics_read,statistics_kivelson,Fisher},
one also expects simultaneously ${\rm Sign}_B\simeq 0$ 
(resp. ${\rm Sign}_F\simeq 0$) as the presence of vortices will ``mess up''
the equal sign of the weights of configurations 
with fixed hole positions. From the results 
displayed in Fig.~\ref{OP_Sign}(b) one clearly identifies a clear 
and {\it simultaneous} rapid crossover of ${\rm Sign}_{\rm F}$ 
and ${\rm Sign}_{\rm B}$ signaling a change of statistics of the 
physical holons by binding/unbinding of vortices:
at small (large) kinetic energy/doping, holon behaves as 
fermions (bosons) as argued from
previous analysis~\cite{statistics_read,statistics_kivelson}.
The separation line between  these two regions have been estimated and reported in
the phase diagrams of Fig.~\ref{Phase_diag} (c,d) as a thick line. At large kinetic 
energy one expects 
the holons to Bose condensate giving rise to
a charge-2e superfluid similar to the one of Refs.~\cite{Frobenius_doped_QDM,Fisher}
and Fig.~\ref{Phase_diag} (a,b). The dotted lines of Fig.~\ref{Phase_diag} (c,d)
limit the area at small $t$ or $x$ where ${\rm Sign}_{\rm F}=1$ within better than
$5\times 10^{-3}$, possibly connected to the superconduting and/or VBC phase.

{\it Discussion and conclusion}---To summarize, 
a non-Frobenius doped QDM retaining 
the Fermi statistics of the original electrons reveals
a rich phase diagram. Holons doped in the VBC host
are shown to behave as fermions at small enough doping and 
to pair up. This leads to a d-wave superconductor (possibly coexisting
with VBC order) which shows striking similarities with the 2D t--J model.
A key feature that permits the existence of this phase is the
absence of PS in the immediate vicinity
of the Mott insultor, in contrast to the Frobenius 
QDM~\cite{Frobenius_doped_QDM}.
At larger doping, with the disappearance of VBC
order, holons behave as bosons and are expected to Bose 
condensate leading to a charge-2e superfluid~\cite{Frobenius_doped_QDM,Fisher}.
In this "statistics transmutation" the role of 
vortices (more specifically $Z_2$ vortices as described 
in Ref.~\cite{Fisher}) is outlined.
A complex behavior is also encountered in-between characterized by a rapid cross-over region where
${\rm Sign}_{\rm F}\sim{\rm Sign}_{\rm B}$ and some tendency towards PS. 
Identifying more precisely this cross-over is beyond the power of present-day computers but various hypothesis can be formulated.
One scenario is a region of PS (i.e. {\it macroscopic} coexistence) of the two limiting 
phases which have been clearly identified 
in this work, as suggested by a slight negative 
curvature of $e(x)$ vs $x$ in Fig.~\ref{Maxwell_bosons}.
Another alternative
is that both (charged) fermionic and bosonic 
excitations coexist at intermediate doping 
in a way which might bear some resemblance with the
``holon-hole superconductor'' of Ref.~\cite{alg_charge_liquids} with, w.r.t. the
large-$x$ superfluid, additional 
low-energy excitations from paired holes. 
Incidentally, the observation of quantum oscillations attributed
to some Fermi surface has been reported recently
in cuprate superconductors~\cite{oscillations}. 
In any case, the tendency towards phase separation
leaves room for modulated/unidirectional structures (or stripes)
once long-range Coulomb repulsion is included.
This is of central interest since modulated structures
have been seen in cuprate superconductors, e.g. in recent 
STM experiments~\cite{STM}.
 
 Note that the present study has been realized on the square lattice where the 
 undoped QDM has VBC order. It would also be of great interest to
 investigate the case of the triangular lattice where 
 a deconfined RVB GS can be realized in the Mott insulator~\cite{triangular}. 
 
% Acknowledgements    
\begin{acknowledgments}
I acknowledge hospitality from the Kavli Institute for 
Theoretical Physics while part of this work was done. 
This research was supported in part by the French Agence Nationale de la 
Recherche (ANR) 
under grant No.~ANR-05-BLAN-043-01 and by the National Science 
Foundation under grant No.~PHY05-51164.
I also thank L.~Balents, M.~Fisher, S.~Sachdev 
and T.~Senthil for enlightening discussions. 
\end{acknowledgments}
% 
% Bibliography        


\begin{thebibliography}{99}
\bibitem{note_ED}  All space group symmetries are used to block-diagonalize the hamitonian.
\bibitem{note_Nh_eq2} For $N_h=2$ and $8\times 8$, there are $102 \, 969 \, 156$ symmetrized states in the GS sector. The GS energy is $-22.336\, 584$ for $V=0$ and $|J|=|t|=1$.
\bibitem{note_equivalence} The equivalence of the two representations has been checked numerically.
\end{thebibliography}
\end{document}